\documentclass[a4paper,11pt]{article}

\usepackage{jinstpub} 

\usepackage{epsfig}
\usepackage{epstopdf}
\usepackage{color} 
\usepackage{url} 

\usepackage{lineno}

\title{Twelve-crystal prototype of Li$_2$MoO$_4$ scintillating bolometers for CUPID and CROSS experiments}

\collaboration{%
\includegraphics[height=17mm]{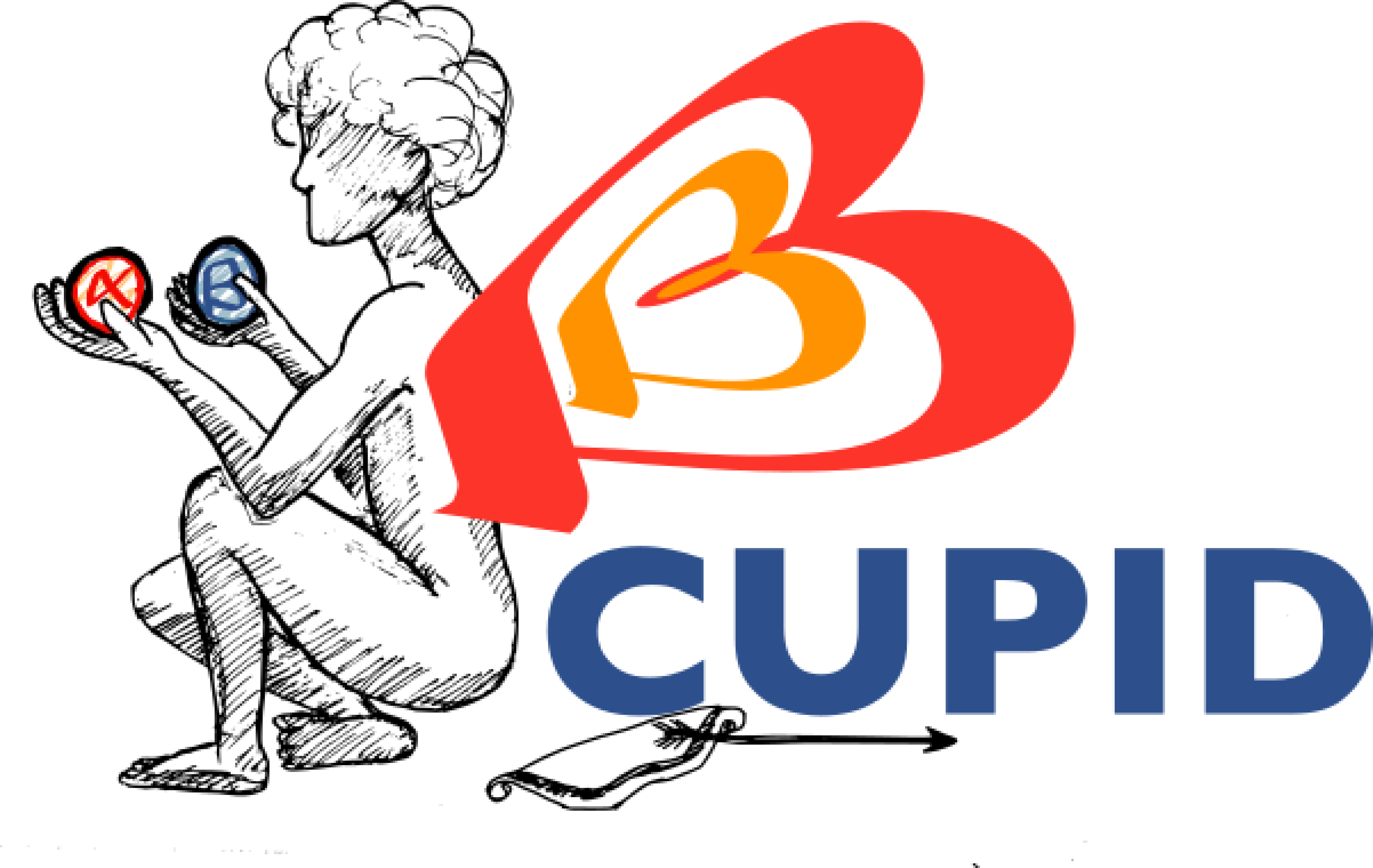}
\includegraphics[height=17mm]{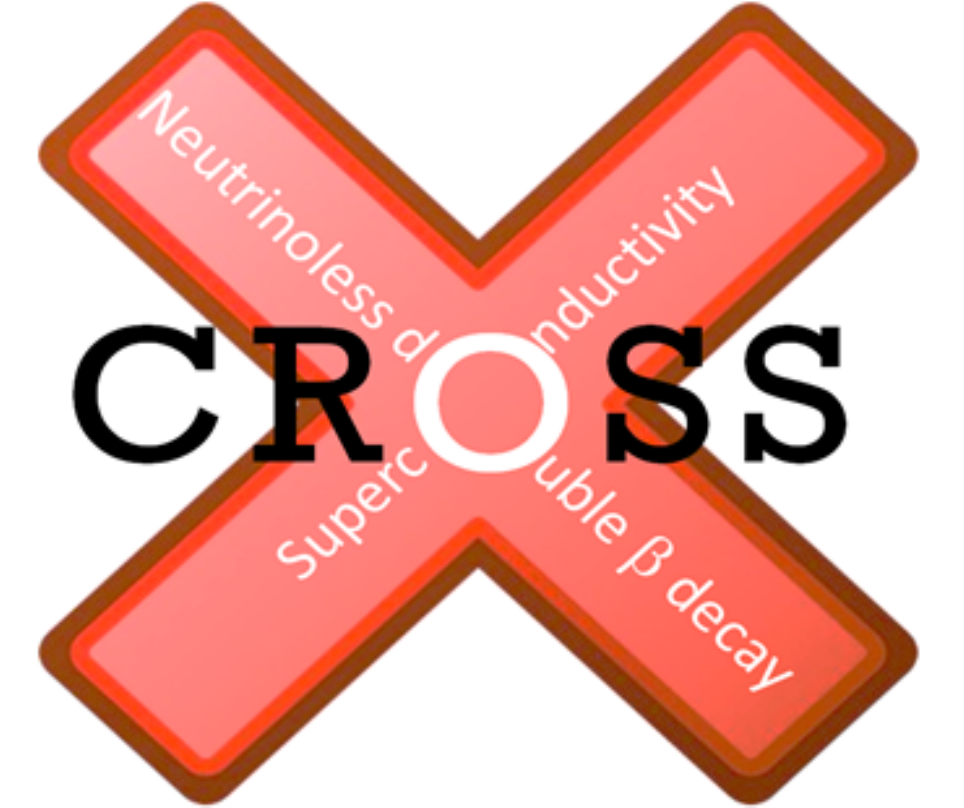}\\[6pt]
CUPID and CROSS collaborations}

\author[a]{K.~Alfonso,}
\author[b]{A.~Armatol,}
\author[c]{C.~Augier,}
\author[d]{F.T.~Avignone~III,}
\author[e]{O.~Azzolini,}
\author[f]{M.~Balata,}
\author[g]{I.C.~Bandac,}
\author[h]{A.S.~Barabash,}
\author[i]{G.~Bari,}
\author[j,k]{A.~Barresi,}
% ...

\affiliation[a]{Virginia Polytechnic Institute and State University, Blacksburg, VA, USA}
\affiliation[b]{IRFU, CEA, Université Paris-Saclay, Saclay, France}
\affiliation[c]{Univ Lyon, Université Lyon 1, CNRS/IN2P3, IP2I-Lyon, Villeurbanne, France}
\affiliation[d]{University of South Carolina, Columbia, SC, USA}
\affiliation[e]{INFN Laboratori Nazionali di Legnaro, Legnaro, Italy}
\affiliation[f]{INFN Laboratori Nazionali del Gran Sasso, Assergi, AQ, Italy}
\affiliation[g]{Laboratorio Subterr\'aneo de Canfranc, Canfranc-Estaci\'on, Spain}
\affiliation[h]{National Research Center Kurchatov Institute, Kurchatov Complex of Theoretical and Experimental Physics, Moscow, Russia}
\affiliation[i]{INFN Sezione di Bologna, Bologna, Italy}
\affiliation[j]{INFN Sezione di Milano-Bicocca, Milan, Italy}
\affiliation[k]{University of Milano-Bicocca, Milan, Italy}
% ...

\emailAdd{cupid.publications@lngs.infn.it}
\emailAdd{cross.publications@lsc-canfranc.es}

\abstract{An array of twelve 0.28 kg lithium molybdate (LMO) low-temperature bolometers equipped with 16 bolometric Ge light detectors, aiming at optimization of detector structure for CROSS and CUPID double-beta decay experiments, was constructed and tested in a low-background pulse-tube-based cryostat at the Canfranc underground laboratory in Spain. Performance of the scintillating bolometers was studied depending on the size of phonon NTD-Ge sensors glued to both LMO and Ge absorbers, shape of the Ge light detectors (circular vs. square, from two suppliers), in different light collection conditions (with and without reflector, with aluminum coated LMO crystal surface). The scintillating bolometer array was operated over 8 months in the low-background conditions that allowed to probe a very low, $\mu$Bq/kg, level of the LMO crystals radioactive contamination by $^{228}$Th and $^{226}$Ra.}


\begin{document}
\maketitle
\flushbottom


\section{Introduction}
\label{sec:intro}

Neutrinoless double-beta ($0\nu\beta\beta$) decay is a hypothetical spontaneous nuclear disintegration, which transforms quasi-simultaneously two neutrons to two protons with emission of only two electrons  \cite{Agostini:2022,Workman:2022,Dolinski:2019,DellOro:2016tmg,Vergados:2016,Pas:2015,Vergados:2012,Rodejohann:2011}. 
In contrast to two-neutrino double-beta ($2\nu\beta\beta$) decay, which results to the emission of additional anti-neutrinos accompanied to two electrons \cite{Pritychenko:2023,Barabash:2020,Belli:2020a,Saakyan:2013}, the $0\nu\beta\beta$ process violates the total lepton number by two units that requires a new physics beyond the Standard Model of particle physics to incorporate the lepton number violation \cite{Agostini:2022,Workman:2022,Dolinski:2019,DellOro:2016tmg,Vergados:2016,Pas:2015,Vergados:2012,Rodejohann:2011}. The existence of $0\nu\beta\beta$ decay is additionally motivated by the discovery of the finite neutrino mass \cite{Workman:2022}, a necessary condition for $0\nu\beta\beta$ process to occur \cite{Schechter:1982,Duerr:2011}. The $0\nu\beta\beta$ decay rate, being proportional to the second power of the lepton number-violating parameter (i.e. an effective Majorana neutrino mass for the ``standard'' $0\nu\beta\beta$ mechanism of light Majorana neutrino exchange), is expected to be significantly reduced in comparison to the $2\nu\beta\beta$ process, already observed in dozen isotopes with the half-lives $\sim$10$^{18}$--10$^{24}$~yr \cite{Pritychenko:2023,Barabash:2020,Belli:2020a}. The most sensitive experiments have recently probed the existence of a $0\nu\beta\beta$ process with half-lives $\sim$10$^{24}$--10$^{26}$~yr for a few $\beta\beta$ isotopes  \cite{Arnold:2015,Anton:2019,Adams:2020,Agostini:2020,Azzolini:2022,Augier:2022,Arnquist:2023,Abe:2023}. To further improve the experimental sensitivity one should increase the isotope of interest exposure, achieve as high as possible energy resolution and low, ideally zero background \cite{Agostini:2022,Adams:2022b,DAndrea:2021,Biassoni:2020,Shimizu:2019,GomezCadenas:2019}.

Bolometers (also called as low-temperature calorimeters) are among the widely and long-term-used technologies in $0\nu\beta\beta$ decay search \cite{Fiorini:1984,Pirro:2005ar,Pirro:2017,Poda:2017,Bellini:2018,Biassoni:2020,Poda:2021,Zolotarova:2021a}, in particular for $^{82}$Se \cite{Azzolini:2022}, $^{100}$Mo \cite{Augier:2022,Kim:2023}, and $^{130}$Te \cite{Brofferio:2018}. Such solid-state detectors work at temperatures close to the absolute zero and register particle interactions using phonon sensors, detecting elementary vibrational motions of a crystal lattice induced by the energy release in the detector medium \cite{Enss:2005a,Enss:2005,Enss:2008}. The bolometric technology features 
``universality'', allowing to use different detector materials (i.e. dielectric diamagnetic crystals, semiconductors) containing isotopes of interest and providing excellent performance (e.g. high energy resolution, high detection efficiency, low energy threshold, particle identification capability) \cite{Pirro:2005ar,Poda:2017,Poda:2021}. 

The largest bolometric $0\nu\beta\beta$ experiment CUORE (Cryogenic Underground Observatory of Rare Events) is currently operating an array of 988 5-cm-cube TeO$_2$ detectors with a total mass of 741 kg at the Gran Sasso National Laboratory (LNGS) in Italy, already acquiring ton$\times$yr exposure \cite{Adams:2022,Adams:2022a}. Despite of a great achievement on realization of experiment using high-performance radiopure detectors operated in very low-background conditions, the lack of particle identification between $\alpha$ and $\gamma$($\beta$) results to the dominant $\alpha$ counting rate of $\sim$10$^{-2}$~counts/yr/kg/keV in the vicinity of $^{130}$Te Q$_{\beta\beta}$-value (2528 keV) \cite{Alduino:2017}, which limits the experimental $0\nu\beta\beta$ sensitivity to $\sim$10$^{26}$ yr \cite{Alduino:2017pni}. 

To improve the sensitivity by an order of magnitude (10$^{27}$ yr) exploiting the existing CUORE infrastructure, a next-generation $0\nu\beta\beta$ bolometric experiment CUPID (CUORE Upgrade with Particle IDentification) has been proposed \cite{Wang:2015raa,CUPIDInterestGroup:2019inu}. $^{100}$Mo-enriched lithium molybdate (Li$_{2}${}$^{100}$MoO$_4$) scintillating bolometers with particle identification capabilities are selected \cite{CUPIDInterestGroup:2019inu} to achieve a background index of $\sim$10$^{-4}$ counts/yr/kg/keV at the Q$_{\beta\beta}$-value of $^{100}$Mo (3034 keV).

The technology of Li$_{2}${}$^{100}$MoO$_4$ low-temperature detectors with phonon-scintillation dual readout, using complementary thin Ge bolometers for photon detection, has been proposed in \cite{Barinova:2010,Cardani:2013}, developed by the LUMINEU  \cite{Armengaud:2017,Grigorieva:2017,Poda:2017a} and validated on a larger scale by the CUPID-Mo  \cite{Armengaud:2020a,Schmidt:2020,Poda:2020} collaborations. In both LUMINEU  and  CUPID-Mo experiments ---realized respectively with 4-crystal \cite{Armengaud:2017,Poda:2017a} and 20-crystal \cite{Armengaud:2020a} arrays operated in the EDELWEISS setup at the Modane underground laboratory in France--- cylindrical crystals ($\oslash$44$\times$45 mm, $\sim$0.21 kg, $^{100}$Mo enrichment $\sim$ 97\%) have been used. 

In order to construct a more compact detector structure to improve multi-site events detection and better occupy the available experimental volume, a cubic shape is preferred. Cubic crystals have been selected for CUPID  \cite{CUPIDInterestGroup:2019inu} as well for CROSS (Cryogenic Rare-event Observatory with Surface Sensitivity), a project to demonstrate the potential of metal-coated bolometers capable of particle identification of near surface interactions \cite{Bandac:2020,Khalife:2020,Khalife:2020a,Zolotarova:2020,Bandac:2021,Khalife:2021}. Several cubic lithium molybdate samples with a size of 45$\times$45$\times$45 mm each ---produced following the LUMINEU crystals fabrication protocol \cite{Berge:2014,Grigorieva:2017,Armengaud:2017}---  have been recently operated by CROSS and CUPID collaborations at the Canfranc underground laboratory (LSC) in Spain \cite{Armatol:2021b}, and at the LNGS \cite{Armatol:2021a,Alfonso:2022}. The results of these tests confirm LUMINEU and CUPID-Mo findings on high energy resolution, high crystal radiopurity, and an efficient scintillation-assisted particle identification despite the modest light output of the material \cite{Armengaud:2017,Armengaud:2020a,Poda:2021}. 

In continuation of a joint CUPID and CROSS R\&D activity aiming at the optimization of the detector structure with cubic crystals, we constructed a 12-crystal array to explore the Li$_{2}${}$^{100}$MoO$_4$ bolometric and scintillation performance using different-size phonon sensors and photodetectors, and operating in different light collection conditions, i.e. with and without reflector. A description of the 12-crystal array, its underground operation at LSC, and the results of detector characterization are subjects of the present paper.


\section{Twelve-crystal array of Li$_2$MoO$_4$ scintillating bolometers}

Twelve cubic-shaped Li$_2$MoO$_4$ (LMO) crystal scintillators with 45-mm side and 0.28 kg each were used for the detector construction. All the crystals are enriched in $^{100}$Mo ($\sim$98\%) \cite{Bandac:2020,Armatol:2021b}, except the one, which is depleted in $^{100}$Mo \cite{Grigorieva:2020}. We randomly chose 11 LMOs from a batch of 32 identical crystals produced for the CROSS experiment \cite{Bandac:2020}. Four crystals were coated with Al film of 2.0~$\mu$m or 0.2~$\mu$m thickness and with/without a SiO underlayer deposited by evaporation on the four lateral faces. The lateral side of another four LMOs was surrounded by a reflective film (Vikuiti{\texttrademark}), while the rest LMOs were operated without a reflector.

Sixteen thin bolometric light detectors (LDs) made of Ge wafers with electron-grade purity were installed to detect the scintillation light from the LMO crystals. All wafers were supplied by Umicore (Belgium) \cite{Umicore}, but two of them (labeled here as LD-s2 and LD-s4) were produced by MTI Corporation (U.S.) \cite{MTI}. Half of the Ge wafers have a circular shape (with a size of $\oslash$45$\times$0.18 mm), as used in CUPID-0 \cite{Azzolini:2018tum} and CUPID-Mo \cite{Armengaud:2020a}. The other wafers are square-shaped (with a size of 45$\times$45$\times$0.30 mm for Umicore and 45$\times$45$\times$0.50 mm for MTI produced slabs) in order to better fit to the shape of LMOs and  thus increase the light collection. All Ge slabs were SiO-coated from both sides (with an effective coating thickness of $\sim70$~nm) to reduce the light reflection \cite{Mancuso:2014,Azzolini:2018tum,Armengaud:2020a}. In addition, a single circular LD has Al concentric electrodes deposited on one surface, which can be used to amplify thermal signals exploiting the Neganov-Trofimov-Luke effect \cite{Novati:2019} (not exploited in the present work).

\begin{figure}
\centering
\includegraphics[width=1.0\textwidth]{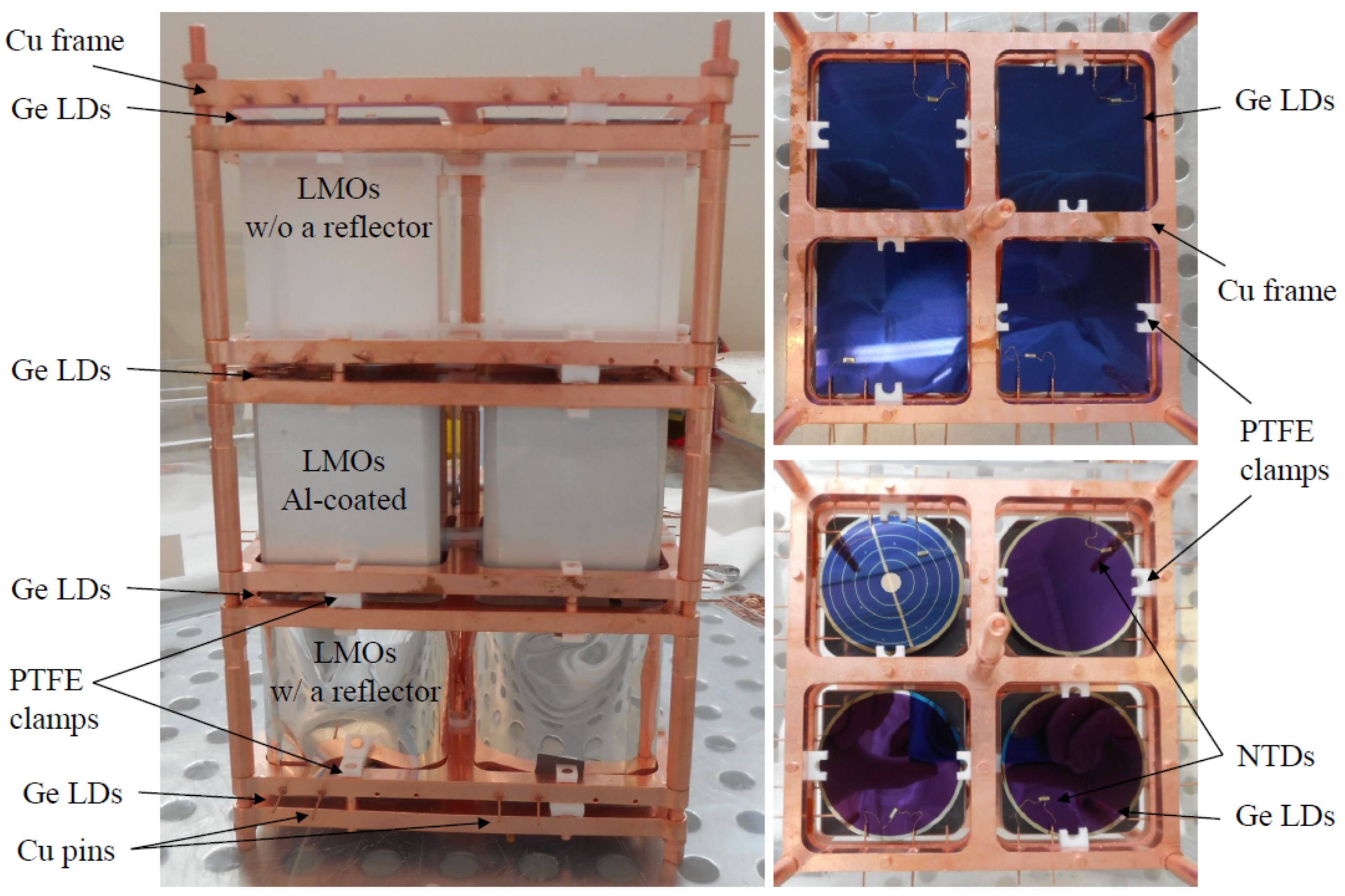}
\caption{(Left) Twelve-crystal array of LMO scintillating bolometers, constructed from three 4-crystal floors containing 45-mm-side cubic LMO scintillating bolometers and four floors Ge wafers acting as light detectors from top and bottom of each LMO in the tower. The top layer of LMO crystals is fully open (no reflector), the middle layer contains only Al-coated LMOs, while the bottom layer LMOs are surrounded by a reflective film. (Right, top) A view on a layer of square-shaped Ge LDs. (Right, bottom) A top view of a layer of circular Ge LDs; one device has concentric Al electrodes to exploit signal amplification based on the Neganov-Trofimov-Luke effect.}
\label{fig:Detector_Photo}
\end{figure}

\nopagebreak

LMO crystals and Ge slabs were equipped with Neutron Transmutation Doped Ge \cite{Haller:1994} thermistors (NTDs). Eight LMOs were instrumented with ``standard''-size NTDs (3$\times$3$\times$1~mm), while larger NTDs (9$\times$3$\times$1~mm) were used for the other LMO crystals to increase phonon collection surface with the aim of improving the signal amplitude. Taking into account a low mass of Ge wafers used in our LDs, these devices were instrumented with small-size sensors to reduce the impact of the sensor's heat capacity: the dimension of twelve of them is 1$\times$3$\times$1~mm, while the rest four are either 0.7$\times$3$\times$1~mm or 0.5$\times$3$\times$1~mm. All NTDs were produced from Ge wafers with similar irradiation parameters yielding uniform thermal responses. For example, the temperature-dependent resistance of NTDs with the sizes of 9$\times$3$\times$1 mm can be approximated as $R(T) = R_0 \cdot e^{(T_0/T )^{0.5}}$ with the parameters $R_0$ $\sim$ 3.6~$\Omega$ and $T_0$ $\sim$ 3.7~K. 
The sensors were epoxy glued onto the crystals surface, using 6 (12) spots of bi-component glue Araldite Rapid for the standard (large) NTDs of LMOs and a drop of glue for the LDs. In addition to NTDs, small heaters (P-doped Si chip \cite{Andreotti:2012}) were glued on the LMOs top with a drop of the Araldite glue and used for the periodical injection of the constant Jules power for the stabilization of the thermal gain \cite{Alessandrello:1998}.

\begin{table}
 \caption{Composition of the 12-crystal assembly. Light collection conditions are specified as reflector, which includes: 1) absence/presence of a reflector material around LMOs; 2) antireflective coating of LDs. Detector IDs are listed in the last column. Crystal LMO-12 is produced from Mo depleted in $^{100}$Mo, while the other samples are $^{100}$Mo-enriched. Four crystals were coated with an Al layer with a thickness of either 2.0 $\mu$m (LMO-5 and LMO-7) or 0.2 $\mu$m (LMO-6 and LMO-8), but the SiO underlayer was deposited only for half of them (LMO-7 and LMO-8). The major part of LMOs and LDs were equipped with NTDs of 3$\times$3$\times$1 mm and 1$\times$3$\times$1 mm, respectively; the exceptions are indicated as follows: $^a$ --- 9$\times$3$\times$1 mm; $^b$ --- 0.7$\times$3$\times$1 mm; $^c$ --- 0.5$\times$3$\times$1 mm.}
\footnotesize
\begin{center}
\begin{tabular}{c|c|c|llll}
 \hline
Floor & Type & Reflector & \multicolumn{4}{c}{Detector ID} \\
 \hline
Ge LD & square & ~ & LD-s1 & LD-s2 & LD-s3 & LD-s4 \\
~ & ~ & ~ & \multicolumn{4}{c}{~} \\
 \hline
LMO & enriched & absent & LMO-1 & LMO-2 & LMO-3~$^a$ & LMO-4~$^a$ \\
~ & ~ & ~ & \multicolumn{4}{c}{~} \\
 \hline
Ge LD & circular & ~ & LD-c1 & LD-c2 & LD-c3 & LD-c4 \\
~ & ~ & ~ & \multicolumn{4}{c}{~} \\
 \hline
LMO & enriched & Al coating of LMO & LMO-5 & LMO-6 & LMO-7 & LMO-8 \\
~ & ~ & (reflective layer) & \multicolumn{4}{c}{~} \\
 \hline
Ge LD & square & ~ & LD-s5 & LD-s6 & LD-s7 & LD-s8 \\
~ & ~ & ~ & \multicolumn{4}{c}{~} \\
 \hline
LMO & enriched, & Vikuiti film  surrounding LMO & LMO-9~$^a$ & LMO-10 & LMO-11~$^a$ & LMO-12 \\
~ & depleted & (reflective film) & \multicolumn{4}{c}{~} \\
 \hline
Ge LD & circular & ~ & LD-c5~$^c$ & LD-c6~$^b$ & LD-c7~$^c$ & LD-c8~$^b$ \\
~ & ~ & ~ & \multicolumn{4}{c}{~} \\
 \hline
 \end{tabular}
  \label{tab:configuration}
 \end{center}
 \end{table}

\normalsize 

\nopagebreak

The sensor gluing and detector assembly have been realized in a clean room (ISO class 4) of the IJCLab (Orsay, France). The mechanical structure to support detectors is based on the design of a 8-crystal prototype tested at LNGS (Italy) \cite{Armatol:2021a} with modifications aiming to reduce the spacing between crystal and light detectors from 6.5 mm to 0.5 mm. The detector structure consists of a Cu frame, polytetrafluoroethylene (PTFE) elements and 4 LMOs or LDs per floor. Prior the use, all Cu and PTFE parts were cleaned by chemical etching (e.g. see a procedure in \cite{Helis:2021}). The NTDs and heaters were ultrasonically wire-bonded with $\oslash$25-$\mu$m Au wires, which are then crimped to copper pins, while Constantan silk-covered wires were crimped to the other side of the pins, providing electrical contacts. Photos of the assembled 12-crystal array together with examples of circular and square-shaped Ge LDs are shown in figure~\ref{fig:Detector_Photo}; the detector composition is listed in table \ref{tab:configuration}.


\section{Detector operation at LSC}

\begin{figure}[hbt]
\centering
\includegraphics[width=0.75\textwidth]{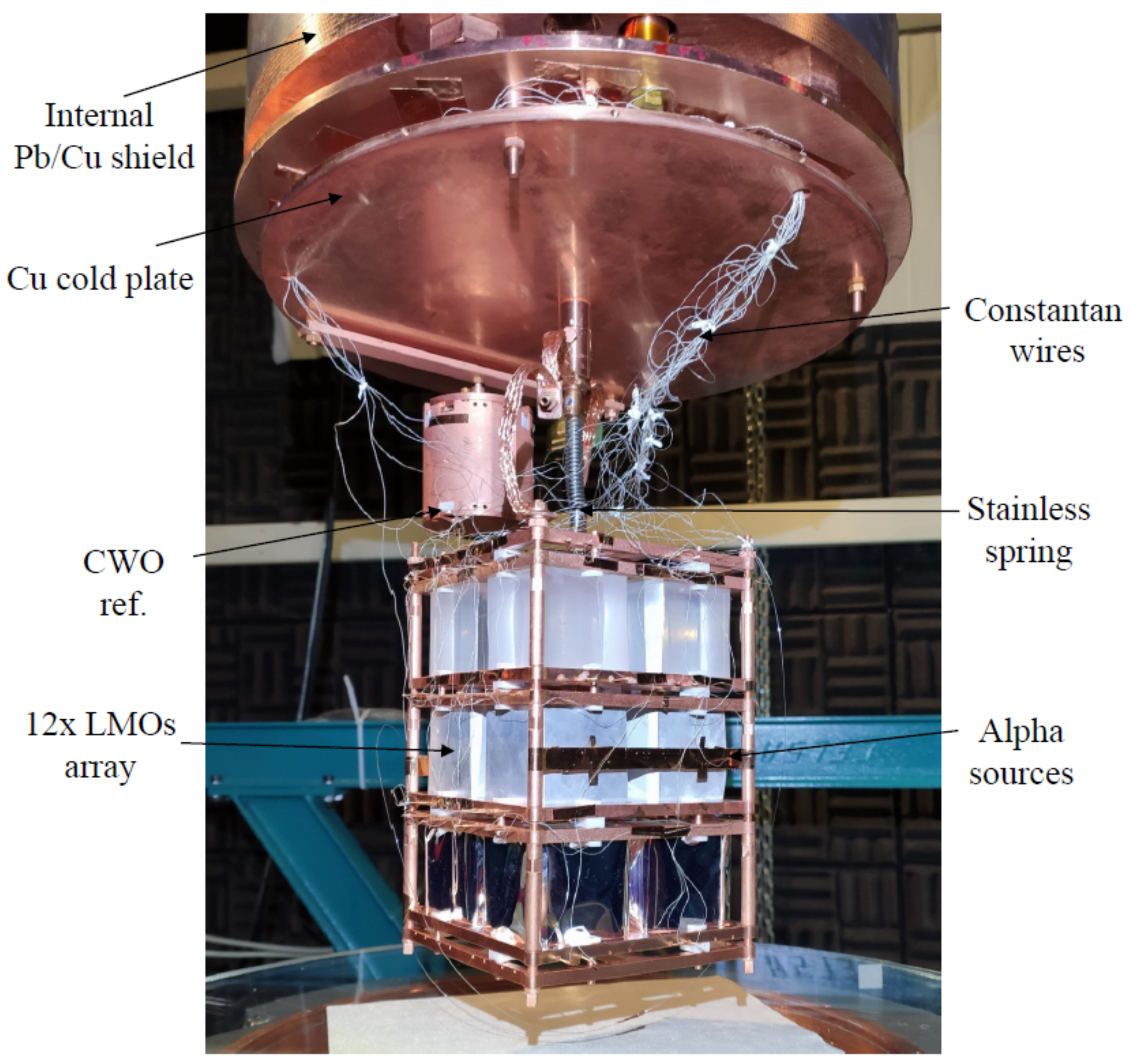}
\caption{The 12-crystal array of LMO scintillating bolometers spring-suspended from the detector plate of the CROSS cryostat. A single module of $^{116}$CdWO$_4$ scintillating bolometer, used as a reference in the set-up \cite{Olivieri:2020,Zolotarova:2020,Helis:2020,Helis:2021}, was mounted directly to the plate to test the noise conditions (not discussed here).  A copper strip in front of the middle layer of LMOs fixes two U alpha sources to probe the particle identification capability of Al-coated LMOs.}
\label{fig:Detector_lsc}
\end{figure}

The tower was transported to LSC and installed in the CROSS cryogenic facility (C2U) \cite{Olivieri:2020,Armatol:2021b}, equipped with a pulse-tube cryostat by CryoConcept (France). 
The dilution refrigerator was cooled down to 4 K by using a pulse-tube Cryomech PT415. In order to mitigate pulse-tube-induced vibrations \cite{Olivieri:2017}, the cryostat is assisted by the Ultra-Quiet Technology{\texttrademark} (UQT) \cite{UQT} to mechanically decouple the pulse tube from the dilution unit. In addition, the 12-crystal array was spring-suspended from the detector plate (see figure \ref{fig:Detector_lsc}) to further reduce vibrational noise.

The underground location of the cryostat allows to have a reduced muon flux thanks to the rock overburden of the laboratory (2450 m water equivalent) \cite{Trzaska:2019}. Additionally, the detector volume inside the cryostat is shielded on the top by a 13 cm thick disk made of interleaved lead and copper (partially seen in figure \ref{fig:Detector_lsc}), and on the sides and bottom by an external 25 cm thick layer of low-radioactivity lead. Finally, a deradonized air flow ($\sim$1 mBq/m$^3$ of Rn \cite{PerezPerez:2022}) around the cryostat had been supplied during the whole time of the experiment. 

The detectors are readout using a low-noise room-temperature DC front-end electronics, restyled from the Cuoricino experiment \cite{Arnaboldi:2002}. The data acquisition (DAQ) system is composed of two 12-channel boards with a programmable 6-pole Bessel-Thomson anti-aliasing filter and integrated 24-bit ADC \cite{Carniti:2020,Carniti:2023}. Since we were able to readout simultaneously only 24 channels from 30 available, we used two configurations of active channels with 6 LDs of the tower switched off (by swapping cables). The injection of thermal pulses through the heaters glued on LMOs is performed using by Keysight 33500B wave-function generator. This generator is also utilized to send periodically a burst of photons, from a room-temperature LED (880 nm) to the experimental volume of the cryostat through an optic fibre, which can be detected by Ge LDs and used for detector optimization and/or stabilization (as heaters for LMOs).

The detectors were maintained at 14 mK for the entire measurement period. All channels, except one LMO with Al coating (LMO-8 in table \ref{tab:configuration}), were found to be operational after the cool down. 
The data taking was realized from the end of September 2020 till the beginning of May 2021 ($\sim$80\% duty cycle over 234 days of measurements). We performed $\gamma$ calibration and background measurements, which amount to 39\% and 58\% of the full data, respectively\footnote{The rest of the data collection was dedicated to neutron calibration (0.5\%) and test measurements (3\%), not described here.}. 
The LMO detectors were calibrated using the most intense $\gamma$ peaks from a $^{232}$Th source made of a thoriated tungsten wire and inserted inside the lead shield. Moreover, this source was also used to calibrate the Ge LDs via the X-ray fluoresence emitted by the LMOs and absorbed by the LDs. Specifically, we foresee detection of a doublet of Mo K$_{\alpha}$ and K$_{\beta}$ X-rays with energies of 17.5 and 19.6 keV, and intensities 65\% and 11\%, respectively. Such method of LD calibration has been recently exploited in \cite{Berge:2018,Poda:2017d,Armengaud:2020a}.

The continuous data were acquired with a sampling rate of 2 kS/s; the cut-off frequency of the low-pass Bessel filter was set to 300 Hz.  The pulser/LED system was used to define the optimal detector working points of LMOs/LDs, optimizing the signal-to-noise ratio \cite{Novati:2019}. All LMOs and a part of LDs were biased with a few nA current on the NTD sensor resulting in a few M$\Omega$ resistance, while another part of LDs were polarized with even higher bias (up to 15 nA) reducing their NTD resistances to hundreds of k$\Omega$.  

The data processing is realized offline using a MATLAB-based analysis tool developed at IJCLab \cite{Mancuso:2016}. The data were triggered and then filtered using the Gatti-Manfredi optimum filter \cite{Gatti:1986}. This matched filter maximizes the signal-to-noise ratio of the filtered data using the information about the signal shape (average signal) and the representative noise spectrum in the frequency domain. For each triggered signal, the program evaluates the signal amplitude (i.e. energy) and calculates several pulse-shape parameters. In order to establish  coincidences between the LMO and LD, we used trigger positions of LMO channels and accounting for a difference in the time response between LMOs and LDs, similarly to the method described in \cite{Piperno:2011}.


\section{Results and discussion}

The results on the 12-crystal prototype characterization are summarized in tables \ref{tab:LMO_performance} and \ref{tab:LD_performance} for LMOs and LDs, respectively. The LMO and LD performance, effect of the light collection conditions and the LMO crystals radiopurity are discussed in the next sections\footnote{Performance of the LMO detector based on $^{100}$Mo-depleted lithium molybdate crystal, LMO-12, is reported here only for the sake of completeness, while a study of such crystals as scintillating bolometers is a subject of a dedicated publication \cite{CROSSdeplLMO:2023}.}.


\begin{table}
 \caption{Performance of LMO bolometers. The detectors instrumented with larger NTDs (9$\times$3$\times$1 mm instead of 3$\times$3$\times$1 mm) are marked with a superscript $^a$. All NTDs were polarized with the 3 nA current. We report the working resistances of NTDs, the rise and decay times, the sensitivity, and the baseline noise resolution of each operational detector. In addition, the energy resolution of the 2615 keV $\gamma$-ray peak of $^{208}$Tl from the $^{232}$Th source, detected by LMOs with a baseline  noise below 10 keV FWHM, is quoted in the last column.}
\footnotesize
\begin{center}
\begin{tabular}{l|c|c|c|c|c|c}
 \hline
LMO ID & $R_{NTD}$ & Rise time & Decay time & Sensitivity & FWHM$_{Noise}$ & FWHM at 2615 keV \\
~ & (M$\Omega$) & (ms) & (ms) & (nV/keV) & (keV) & (keV)  \\
 \hline
LMO-1       & 2.4   & 16    & 111  & 19     & 4.6   & 7.4(4) \\ 
LMO-2       & 2.4   & 24    & 100  & 14     & 36    & -- \\
LMO-3 $^a$  & 4.3   & 11    & 85   & 35     & 26    & -- \\
LMO-4 $^a$  & 3.7   & 14    & 103  & 22     & 5.0   & 7.8(3) \\
 \hline
LMO-5       & 0.90  & 3.9   & 68    & 5     & 93    & -- \\
LMO-6       & 2.2   & 7.0   & 74    & 11    & 16    & -- \\
LMO-7       & 0.54  & 5.5   & 53    & 3     & 27    & -- \\
\hline
LMO-9 $^a$  & 4.8   & 17    & 126   & 32    & 3.8   & 6.9(5) \\
LMO-10      & 2.7   & 23    & 174   & 8     & 8.8   & 11.3(13) \\
LMO-11 $^a$ & 5.3   & 15    & 96    & 40    & 4.4   & 7.6(3) \\
LMO-12      & 3.0   & 16    & 97    & 29    & 3.8   & 6.8(3) \\
\hline
 \end{tabular}
  \label{tab:LMO_performance}
 \end{center}
 \end{table}

\normalsize

\nopagebreak

\begin{table}
 \caption{Performance of LDs. The devices instrumented with smaller NTDs, 0.7$\times$3$\times$1 mm or 0.5$\times$3$\times$1 mm instead of 1$\times$3$\times$1 mm, are marked with superscripts $^b$ and $^c$, respectively. More than a half of LD NTDs were polarized with 15 nA current; the exceptions are LD-s2 (5 nA), LD-s3 (8 nA), LD-s7 (3 nA), and the bottom layer of LDs (2 nA each). We report the NTD resistance at the working point, the rise and decay time parameters, the voltage signal amplitude per unit deposited energy, the baseline noise resolution at the optimum filter output and the resolution for the 17.5 keV X-ray peak of Mo.}
\footnotesize
\begin{center}
\begin{tabular}{l|c|c|c|c|c|c}
 \hline
LD ID & $R_{NTD}$ & Rise time & Decay time & Sensitivity & FWHM$_{Noise}$ & FWHM at 17.5 keV \\
~ & (M$\Omega$) & (ms) & (ms) & ($\mu$V/keV) & (keV) & (keV)  \\
 \hline
LD-s1       & 0.47  & 1.5   & 4.6   & 0.39  & 0.32  & 1.0(3) \\ 
LD-s2       & 0.72  & 1.3   & 5.1   & 0.42  & 0.41  & 1.4(4) \\ 
LD-s3       & 2.2   & 1.7   & 5.3   & 0.36  & 0.32  & 1.2(2) \\ 
LD-s4       & 0.63  & 1.4   & 5.9   & 0.40  & 0.30  & 1.1(1) \\ 
 \hline
LD-c1       & 0.43  & 1.4   & 4.2   & 0.55  & 0.19  & 0.88(7) \\ 
LD-c2       & 0.46  & 1.3   & 3.2   & 0.59  & 0.40  & 1.9(4) \\ 
LD-c3       & 0.40  & 1.3   & 3.7   & 0.80  & 0.18  & 0.95(5) \\ 
LD-c4       & 0.40  & 1.4   & 4.0   & 0.60  & 0.16  & 0.96(5) \\ 
 \hline
LD-s5       & 0.56  & 1.5   & 5.4   & 0.22  & 0.46  & 1.0(1) \\ 
LD-s6       & 0.47  & 1.7   & 5.4   & 0.33  & 0.33  & 1.1(1) \\ 
LD-s7       & 5.6   & 1.7   & 4.2   & 1.2   & 0.19  & 1.0(1) \\ 
LD-s8       & 0.47  & 1.6   & 5.2   & 0.38  & 0.34  & 0.94(6) \\ 
 \hline
LD-c5 $^c$  & 4.1  & 2.0   & 6.7   & 2.0   & 0.09  & 0.64(7) \\ 
LD-c6 $^b$  & 4.9  & 2.0   & 7.1   & 2.7   & 0.06  & 0.88(7) \\ 
LD-c7 $^c$  & 4.2  & 2.0   & 6.1   & 2.4   & 0.08  & 0.83(7) \\ 
LD-c8 $^b$  & 4.4  & 2.0   & 7.8   & 2.2   & 0.06  & 0.90(6) \\ 
 \hline
 \end{tabular}
  \label{tab:LD_performance}
 \end{center}
 \end{table}

\normalsize 


\subsection{Time constants of LMO and LD signals}

Fast time response is essential for LMO scintillating bolometers as a viable tool for the rejection of pile-up events, i.e. a source of a notable background in $^{100}$Mo-enriched thermal detectors for $0\nu\beta\beta$ decay search \cite{Chernyak:2012,Chernyak:2014,Chernyak:2017,Armatol:2021,CROSSpileup:2023}. 
In order to characterize a time response of each triggered signal, we computed the following two parameters: 1) rise time defined as the time required by the signal to increase from 10\% to 90\% of its height; 2) decay time, the time required to decrease from 90\% to 30\% of its height. The rise and decay time parameters for signals of LMO (2--3 MeV events) and LD (12--24 keV) are listed in table \ref{tab:LMO_performance} and \ref{tab:LD_performance}, respectively. 

The time response of the top and bottom LMO floors, as well as all LDs, is found to be similar. Also, we see a hint of a bit faster rise time of LMOs equipped with larger NTDs. Moreover, all operated Al-coated LMOs have two--three times faster signals than typically reported time constants of NTD-instrumented bolometers based on similar-size LMO crystals \cite{Bekker:2016,Armengaud:2017,Armengaud:2020a,Armatol:2021a,Armatol:2021b}. This speed-up of the time response is achieved thanks to the metal coating film, which is responsible for the faster thermalization of phonons induced by a nuclear event \cite{Bandac:2020,Bandac:2021}. 

All LDs equipped with smaller NTDs, being operated at colder working temperatures, are characterized by slightly longer rising part of signals than those of LDs instrumented with larger NTDs being strongly polarized; the LDs time constants and their dependence on working point are within our expectations \cite{Beeman:2013b,Armengaud:2020a,Armatol:2021b}.

\subsection{Performance of LMO bolometers}

\begin{figure}[hbt]
\centering
\includegraphics[width=0.8\textwidth]{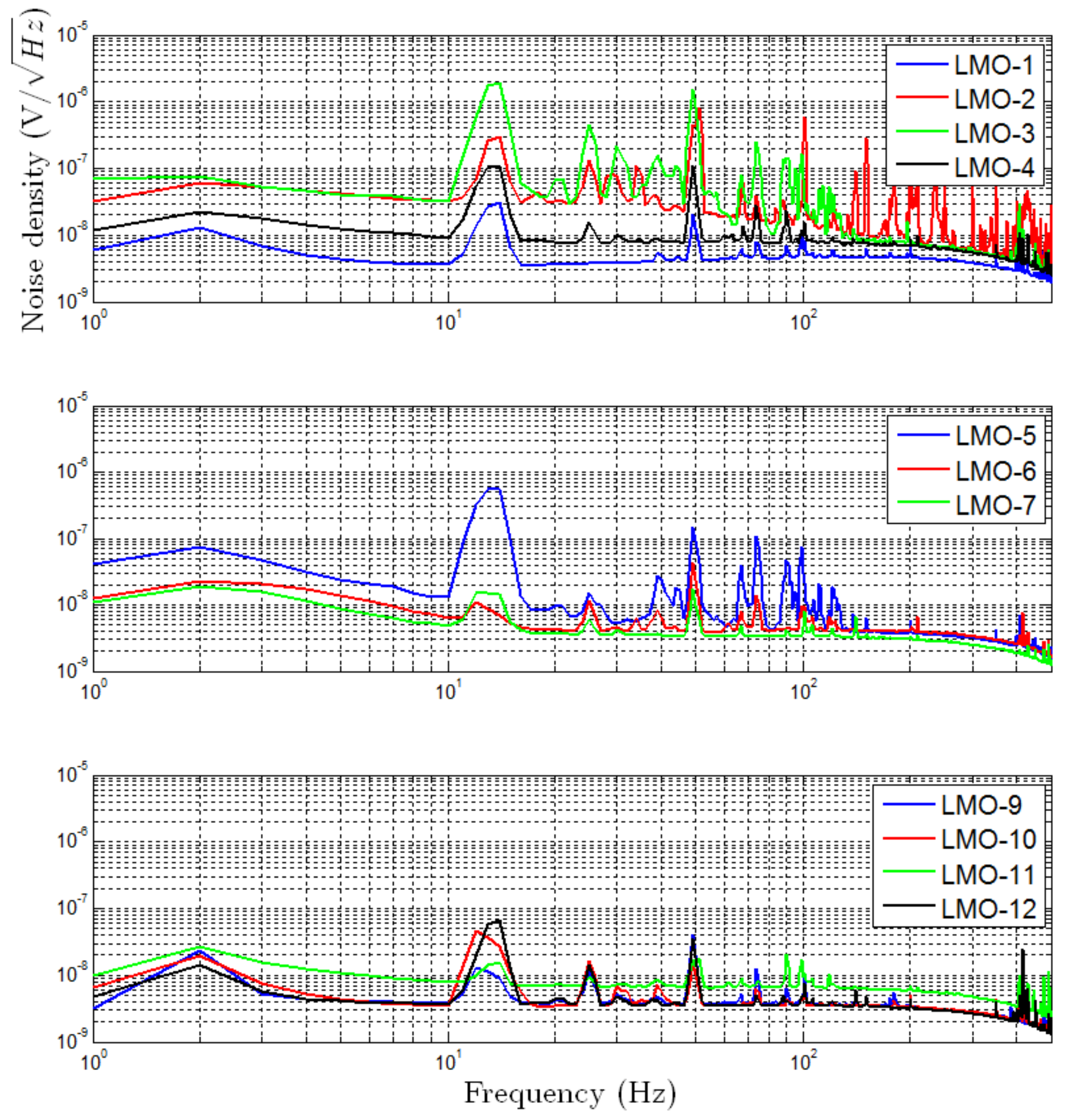}
\caption{Noise density of LMO bolometers operated in the C2U set-up at LSC. The data exhibit a strong 11 Hz peak and other higher frequency harmonics of the main pulse tube frequency at 1.4 Hz. The 50 Hz peak and its harmonics are also present at a notable level for all channels; they are related to electrical configuration external to the set-up. The noise plateau is within the expectations accounting the contributions from the amplifier and the Johnson noise of the load resistors and NTDs. The noise above 300 Hz is reduced by the Bessel filter used in the readout chain.}
\label{fig:LMO_noise}
\end{figure}

The performance of the LMO detectors is found to be reasonably good but not excellent, due to a high microphonic noise level. In particular, for all LMO channels we observed a noise peak at $\sim$11 Hz, in the middle of the signal bandwidth ($\sim$1--50 Hz, taking into account the time constants of LMOs listed in table~\ref{tab:LMO_performance}). The size of this disturbance was channel-dependent, and went 5--2000 nV/$\sqrt{Hz}$ for the chosen working points, as shown in figure~\ref{fig:LMO_noise}. This peak is present even at room temperature when the impedance of the NTDs is negligible. The peak disappears when the pulse-tube is switched off, however the detectors can be operated in such conditions only for a very short time. Also, we measured the vibration profile at 300 K flange using an accelerometer and we found that the pulse tube main frequency at 1.4 Hz is efficiently cut by the UQT system of the cryostat, while harmonics including a prominent resonance close to 11.2 Hz are visible along the vertical and horizontal directions. 
It has to be emphasised that a similar noise structure is also present in the data of the previous cryogenic run \cite{Zolotarova:2020a,Armatol:2021b}, but the $\sim$11 Hz peak is found to be narrow and with the amplitude up to a few tens of nV/$\sqrt{Hz}$, which does not spoil drastically the detector performance. Thus, the difference in the noise impact between two subsequent cryoruns indicates that the suspension used for the 12-crystal array is not optimal and the mechanical decoupling system of the cryostat has to be improved\footnote{Recently, the spring-suspended system has been replaced with the Kevlar wires connected to the three magnetic dampers installed at the still ($\sim$800 mK) stage; the reference detector, a 0.58 kg $^{116}$CdWO$_4$ scintillating bolometer, demonstrated a factor 2 improved noise of both heat and light channels compared to the best previous measurements \cite{Helis:2021} in the C2U set-up.}.

In order to mitigate the noise issue, the detectors were polarized with high currents (3 nA), not optimal in terms of the detector sensitivity. Consequently, the voltage signal amplitudes of LMOs ($\sim$10--40 nV/keV, see table \ref{tab:LMO_performance}) are comparatively low for the 14 mK temperature. Indeed, these results are similar to the sensitivity of LMOs in CUPID-Mo operated at 21 mK temperature \cite{Armengaud:2020a}. It is worth noting a higher sensitivity of LMOs instrumented with larger NTDs, in correlation with the larger resistance values measured at the same applied current. Thus, we can conclude that the gain in the sensitivity for these LMOs is not due to the improved coupling between the sensor and the absorber, but is mostly an effect of the NTD resistance.

All three operational LMOs with Al-coated lateral surfaces show that substantial coverage of the detector surface with Al film leads to sensitivity reduction, as observed e.g. in the CROSS R\&D program \cite{Bandac:2021,Khalife:2021}. We also confirm this negative effect on detector performance. The sensitivities of the three surviving Al-coated LMO detectors are reduced by more than a factor 2 on average with respect to uncoated crystals.

\begin{figure}[hbt]
\centering
\includegraphics[width=0.6\textwidth]{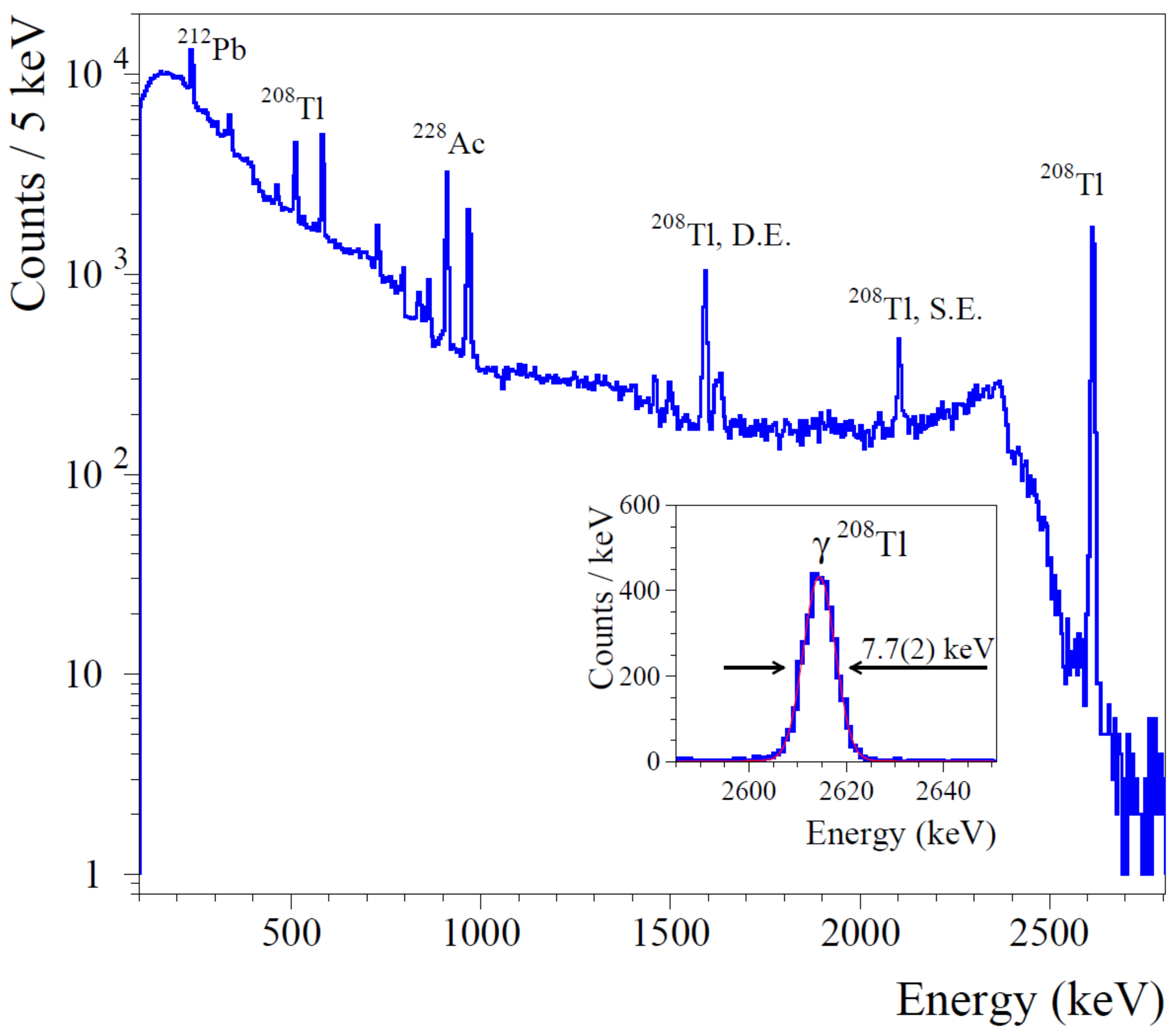}
\caption{Energy spectrum of a $^{232}$Th source measured with the 5 best-performing $^{100}$Mo-enriched LMOs. The most intense $\gamma$-ray peaks observed in the spectrum are labeled (D.E. and S.E. are double and single escape peaks, respectively). A fit to the 2615 keV peak of $^{208}$Tl is shown in the inset; the energy resolution is 7.7(2) keV FWHM.}
  \label{fig:LMO_Th_spectrum}
\end{figure}

The energy resolution of the baseline noise after the optimum filtering is reasonably good for all uncoated LMOs, $\sim$4--5 keV FWHM, but two detectors characterised by tens keV FWHM noise, similar to Al-coated LMOs. Despite the noise problems, most modules of the top and bottom LMO floors exhibit a good energy resolution at high energy $\gamma$ quanta (2.6 MeV), $\sim$7--8 keV FWHM, close to the demands of $\sim$5 keV FWHM for CUPID \cite{CUPIDInterestGroup:2019inu,Wang:2015raa} and CROSS \cite{Bandac:2020}. 
Despite some difference in detectors' sensitivity, we did not observe a difference in energy resolution between LMOs instrumented with standard size NTDs and the larger ones. 
Also, it is interesting to note that the resolution of the 2615 keV $\gamma$ peak is much better than one can expect from the baseline noise values; typically a factor 2--3 difference is reported \cite{Armengaud:2017,Armengaud:2020a,Armatol:2021a,Armatol:2021b,Alfonso:2022}. Thus, a large part of the measured noise does not have a notable impact on the variation of the LMO thermal signal.

A good energy resolution of LMO bolometers can be seen in figure \ref{fig:LMO_Th_spectrum}, where the combined calibration spectrum measured with the five best-performing $^{100}$Mo-enriched LMOs is shown. In particular, the energy resolution of a 2615 keV $\gamma$ peak in the combined spectrum is 7.7(2) keV FWHM. The energy resolution of well-performing LMOs is similar to the performance of the predecessor 8-crystal tower tested in the LHe-bath cryostat at LNGS \cite{Armatol:2021a} and of the 20-crystal array of CUPID-Mo operated at LSM in the  EDELWEISS set-up \cite{Armengaud:2020a}), which utilises both the LHe-bath and thermal machines \cite{Armengaud:2017b}.

\subsection{Performance of Ge light detectors}

\begin{figure}[hbt]
\centering
\includegraphics[width=0.8\textwidth]{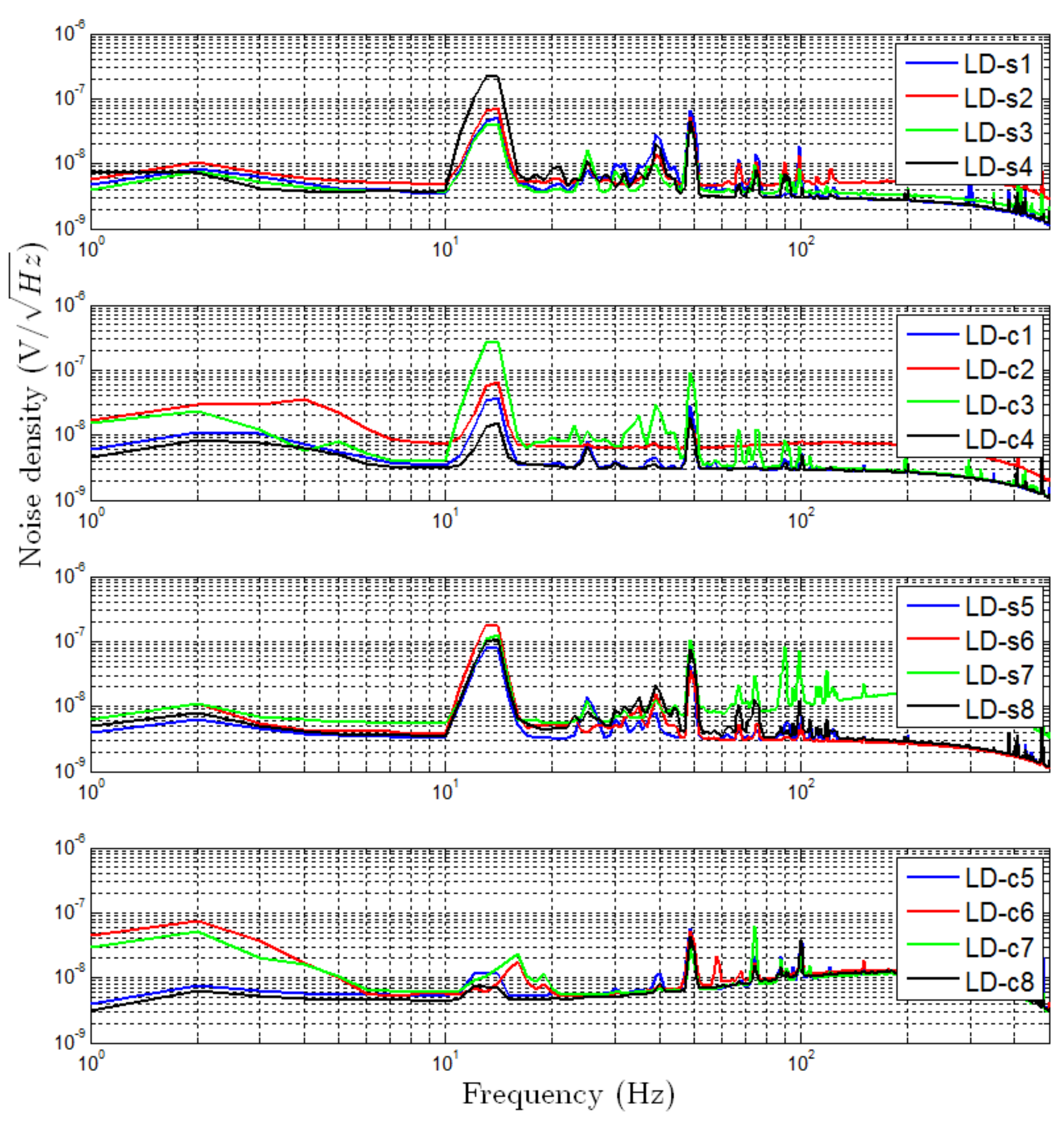}
\caption{Noise density of bolometric Ge LDs operated in the C2U set-up at LSC. The LD noise spectra are similar to the data of LMOs (figure~\ref{fig:LMO_noise}).}
\label{fig:LD_noise}
\end{figure}

The noise power spectra of the LDs also exhibit the 11 Hz peak with an amplitude $\sim$3--300 nV/$\sqrt{Hz}$, as illustrated in figure \ref{fig:LD_noise}. The response of LDs is faster than LMOs, thus a high frequency noise up to hundred(s) Hz lies in the signal bandwidth, affecting the LD performance.

The least noise affected LDs are the four devices in the bottom floor of the tower. This behaviour is surprising, given that all detectors were assembled in the same way. Consequently, these four LDs exhibit the highest sensitivity in terms of the signal amplitude (2.0--2.7 $\mu$V/keV) and a very low noise level (60--90 eV FWHM), as seen in table \ref{tab:LD_performance}. Such good performance is similar to the results of circular LDs tested in the previous cryogenic run of the C2U set-up \cite{Zolotarova:2020a,Armatol:2021b}, with a minor impact of the 11 Hz noise. 

Other LDs were operated with stronger currents across NTDs, thus their signal amplitudes are significantly reduced (mostly to 0.2--0.6 $\mu$V/keV). Using the data collected for the working points scanning with the LED system (i.e. the amplitude of the LED-induced signal as a function of the NTD current), we estimated the reduction factor to compare the LDs sensitivity at the same NTD current. We found the most LDs being polarized at 2 nA (as the bottom layer of LDs) should have sensitivities in the range of 1--3 $\mu$V/keV, i.e. similar to the best-performing LDs of the present work. 

By comparing the detectors operated at similar working points we see that the circular LDs mostly outperform the squared ones in terms of the sensitivity and the baseline energy resolution; a factor of 2 difference in the noise resolution can be attributed to the difference in the LDs sensitivity. The gain in the signal amplitude of the circular LDs can be explained by a reduced heat capacity of these devices, taking into account a factor 2--4 smaller volumes than those of the square-shaped LDs. However, we found that the noise resolution of the square-shaped LDs is slightly worse ($\sim$7\% in average) than expected from the drop of their sensitivity, which might indicate another sources of the noise degradation, e.g. a bit higher impact of vibrations on these devices compared to the circular LDs. 

\begin{figure}[hbt]
\centering
\includegraphics[width=0.49\textwidth]{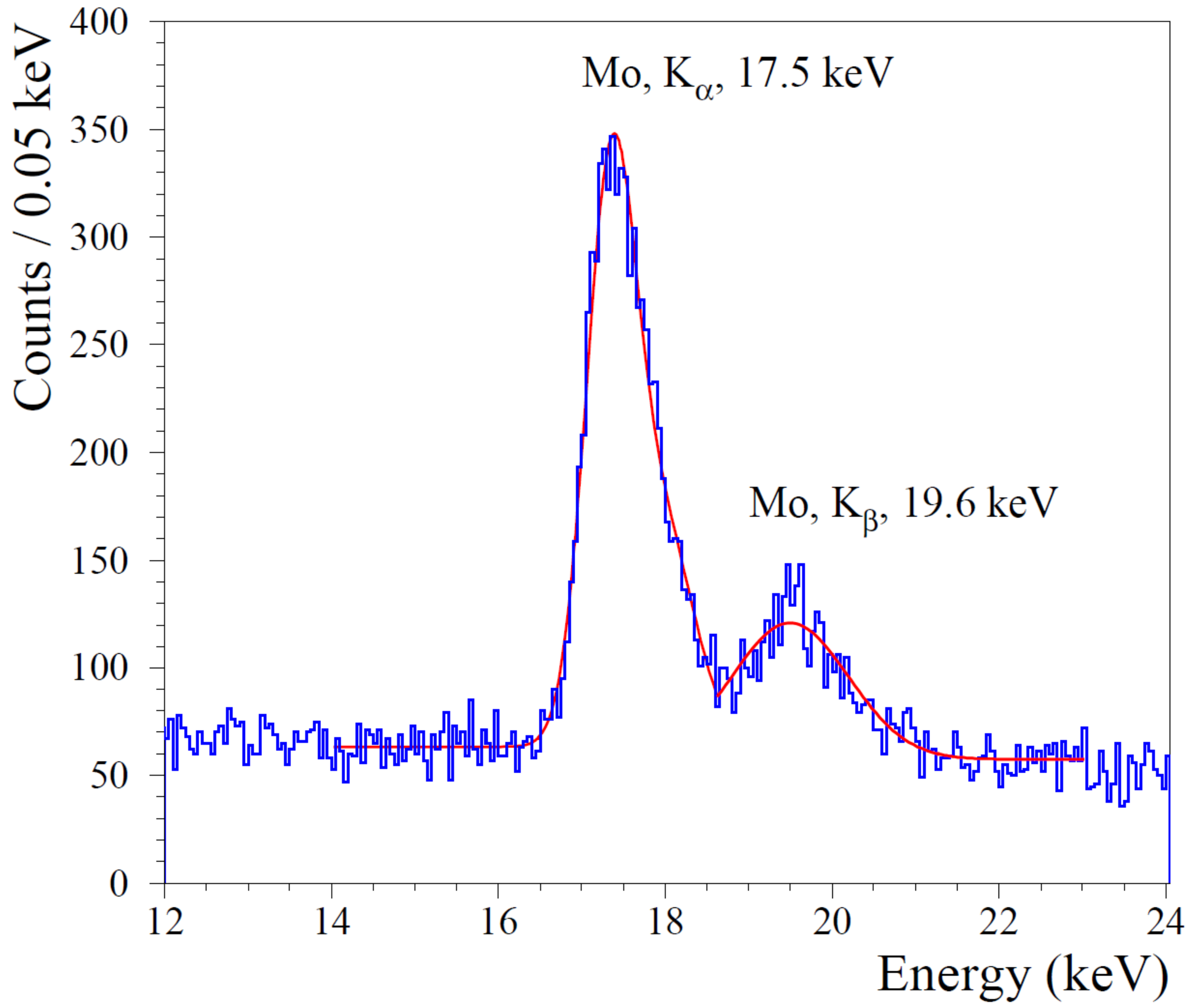}
\caption{Energy spectrum of Mo X-rays detected by all Ge LDs (except 3, rejected due to a poor performance and low statistics of Mo X-rays) in the 266-h-long measurements with a ``high-intensity'' Th source. The fit to the spectrum with a linear function plus two Gaussians is shown in a solid red line; the energy resolution of the 17.5 keV X-ray peak in the combined spectrum is 0.72(2) keV FWHM.}
  \label{fig:LD_Mo_spectrum}
\end{figure}

Irrespective of a large difference in the baseline noise of LDs (up to a factor of 7), the energy resolution for the Mo X-rays is found to be rather similar. We explain this effect by a position dependence of the detector response on the impact point, which is typical for small and thin absorbers \cite{Armengaud:2017}. The combined energy spectrum accumulated by all Ge LDs except three of them (LD-s2, LD-s4, and LD-c2 characterized by high noise and low statistics of the Mo doublet) is presented in figure~\ref{fig:LD_Mo_spectrum}.

\subsection{Scintillation signal in different light collection conditions}

\begin{figure}[hbt]
\centering
\includegraphics[width=0.7\textwidth]{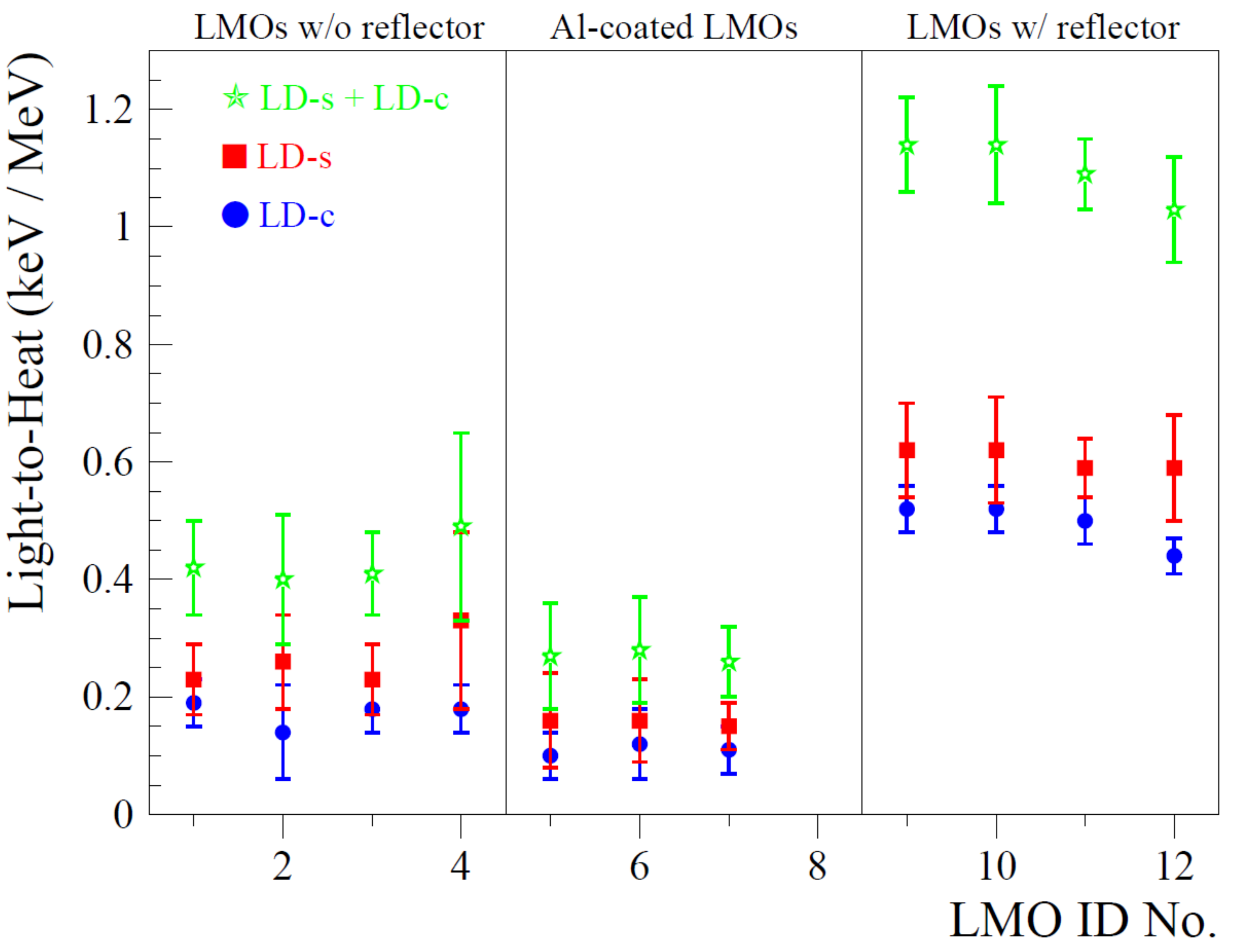}
\caption{Light-to-heat ratio for $\gamma$($\beta$) events measured by LMO scintillating bolometers in different light collection conditions.}
\label{fig:LMO_LY}
\end{figure}

The 12-crystal tower compostion implies the study of LMO scintillation detection in different light collection conditions, depending on the presence / absence of a reflector around LMO crystal and on the active area (shape) of the LD. 
For the characterization of the light collection conditions we selected $\gamma(\beta)$ events with energies in the 2--3 MeV range and computed for them a light-to-heat parameter (L/H$_{\gamma(\beta)}$), defined as a ratio of a scintillation signal detected by a LD (in keV) per energy deposited as heat in LMO (in MeV). 
Each distribution of the L/H$_{\gamma(\beta)}$ parameter was fitted by a Gaussian function to extract a mean value. Uncertainties of the mean values returned by the Gaussian fits are very small (0.1\%--1.2\%) and do not include systematic effects. Therefore, we use an RMS of each L/H$_{\gamma(\beta)}$ distribution as an uncertainty estimate. Despite the fact that this approach is too conservative and largely overestimates real uncertainties \footnote{By comparing two datasets for LMOs with the reflective film, we see no / minor (1--2)\% difference in the L/H$_{\gamma(\beta)}$ mean values for circular LDs, while the approach based on the L/H$_{\gamma(\beta)}$ width results to the uncertainty estimate on the level of (7--8)\%.}, it has no impact on interpretation of results of the present study. The results on scintillation detection in different light collection condition, expressed by L/H$_{\gamma(\beta)}$, are presented in table \ref{tab:LMO_LY} and illustrated in figure \ref{fig:LMO_LY}.


\begin{table}
 \caption{Light-to-heat energy ratio of $\gamma$($\beta$) events (L/H$_{\gamma(\beta)}$) detected by LMO scintillating bolometers in different light collection conditions, related to a) the absence/presence of a reflector and b) shapes of Ge LD (circular or square) facing LMOs.}
\footnotesize
\begin{center}
\begin{tabular}{c|l|c|c|c|c}
 \hline
Reflector & LMO ID & L/H$_{\gamma(\beta)}$ LD-s & L/H$_{\gamma(\beta)}$ LD-c & L/H$_{\gamma(\beta)}$ sum  & L/H$_{\gamma(\beta)}$ ratio \\
~ & ~ & (keV/MeV) & (keV/MeV) & (keV/MeV)   & LD-s / LD-c \\
 \hline
Absent & LMO-1 & 0.23(6) & 0.19(4) & 0.42(8) & 1.3(5) \\
~      & LMO-2 & 0.26(8) & 0.14(8) & 0.40(11) & 1.9(11) \\
~      & LMO-3 & 0.23(6) & 0.18(4) & 0.41(7) & 1.3(4) \\
~      & LMO-4 & 0.33(15) & 0.18(4) & 0.49(16) & 1.8(9) \\
 \hline
Al-coating of LMO & LMO-5 & 0.16(8) & 0.10(4) & 0.27(9) & 1.6(10)  \\
~                 & LMO-6 & 0.16(7) & 0.12(6) & 0.28(9) & 1.3(8)  \\
~                 & LMO-7 & 0.15(4) & 0.11(4) & 0.26(6) & 1.3(6) \\
\hline
Reflective film & LMO-9  & 0.62(8) & 0.52(4) & 1.14(8)  & 1.2(1) \\
~               & LMO-10 & 0.62(9) & 0.52(4) & 1.14(10) & 1.2(2) \\
~               & LMO-11 & 0.59(5) & 0.50(4) & 1.09(6)  & 1.2(2) \\
~               & LMO-12 & 0.59(9) & 0.44(3) & 1.03(9)  & 1.3(2) \\

\hline
 \end{tabular}
  \label{tab:LMO_LY}
 \end{center}
 \end{table}

\normalsize

In most cases we see $\sim$(20--30)\% higher scintillation signals detected by a squared LD compared to a circular LD faced to the same crystal. This agrees with the expectation, if we consider the 30\% larger area of squared LDs. In a few cases, particularly for MTI Ge based LDs, this difference is about 80\%, which may be explained by more efficient light absorption for these wafers (a similar effect is observed in CUPID-Mo \cite{Armengaud:2020a}).

\begin{figure}[hbt]
\centering
\includegraphics[width=0.49\textwidth]{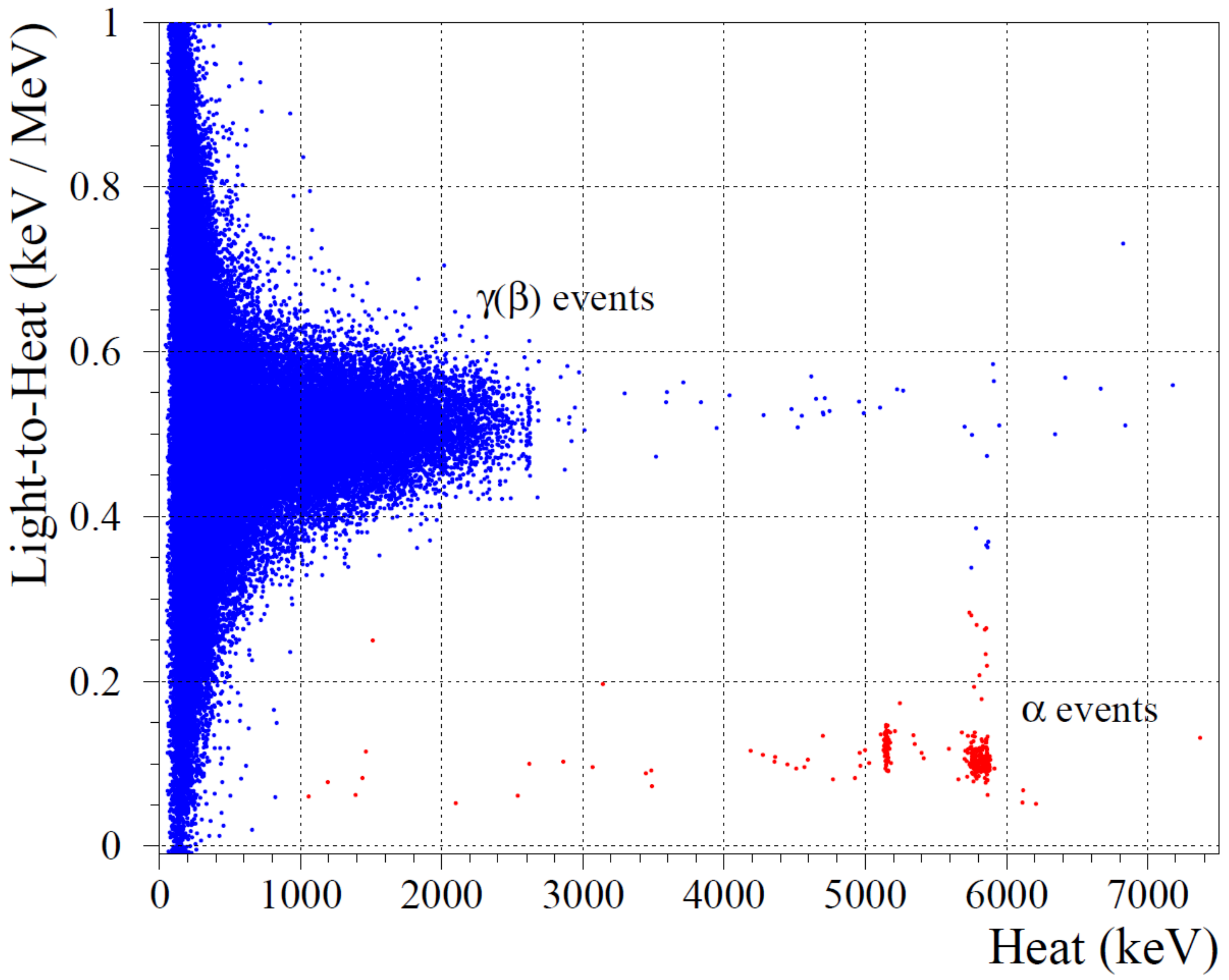}
\includegraphics[width=0.49\textwidth]{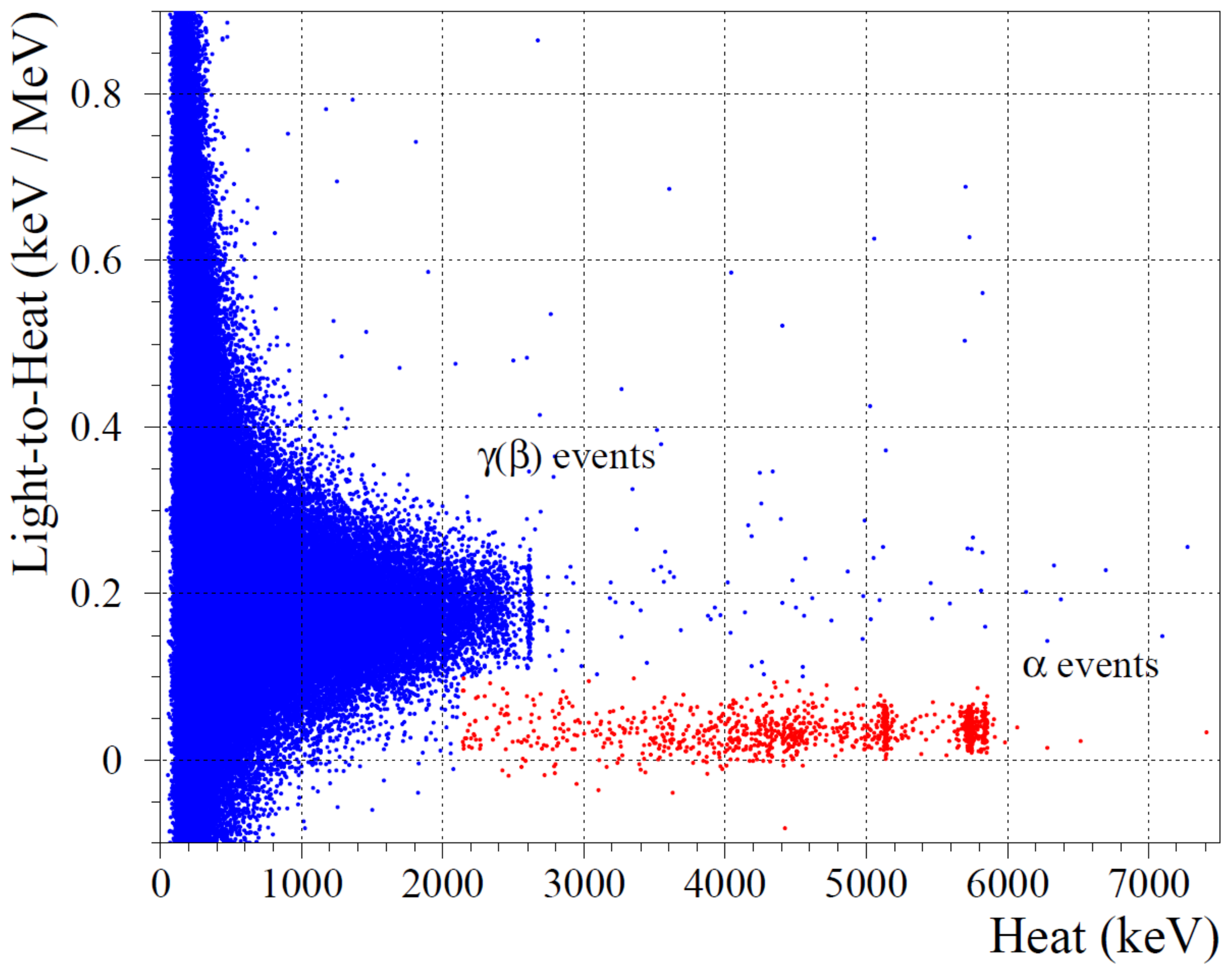}
\caption{Distribution of the light-to-heat parameter as a function of energy for $\gamma$($\beta$) and $\alpha$ events detected by LMOs in different light collection conditions: with a Vikuiti{\texttrademark} reflective film surrounding LMO (left panel) and without a reflector around LMO (right panel). The latter LMO detector was additionally irradiated by the smeared $\alpha$ source (see in text). Circular LDs were utilized in both cases.}
\label{fig:LMO_LY-vs-Heat}
\end{figure}

The largest scintillation signal is detected for crystals surrounded by a reflective film; the weighted average values of the L/H$_{\gamma(\beta)}$ parameters of LMOs coupled to squared and circular LDs are 0.60(4) and 0.49(2) keV/MeV, respectively. An example of the distribution of the L/H parameter is shown in figure \ref{fig:LMO_LY-vs-Heat} (left). Similar light-to-heat ratios were obtained in more recent tests of cubic LMOs with the same size \cite{Armatol:2021a,Armatol:2021b,Alfonso:2022}, and in the LUMINEU and CUPID-Mo experiments realized with cylindrical LMOs \cite{Armengaud:2017,Armengaud:2020a,Poda:2020}.

The absence of a reflector around LMOs drastically affects the photon collection on LDs and thus it leads to substantially reduced L/H$_{\gamma(\beta)}$ parameters of 0.24(4) and 0.18(2) keV/MeV for LMOs viewed by square and circular LDs, respectively. Such situation is illustrated in figure \ref{fig:LMO_LY-vs-Heat} (right). These results are similar to what was observed in the recent tests with the 8-crystal predecessor \cite{Armatol:2021a} and the 4-crystal prototype of the CUPID baseline detector structure \cite{Alfonso:2022}. 

Surprisingly, the worst scintillation collection is observed for all three operational Al-coated LMO bolometers; the average L/H$_{\gamma(\beta)}$ values are 0.15(3) and 0.11(2) keV/MeV for LMOs viewed by square and circular LDs, which is almost twice lower with respect with no coating. Therefore, the present technology of Al-coated bolometers (with a large fraction of the covered surface) developed in CROSS \cite{Bandac:2020,Khalife:2021}, does not satisfy the required performance in terms of the detector sensitivity and the light collection efficiency. We remind that this solution is in principle extremely favorable as it would have allowed surface background rejection by coincidences between nearby crystals. Thus, new investigations to test another technology of Al coating are in progress.

\subsection{Radiopurity of Li$_2$$^{100}$MoO$_4$ crystal scintillators}

\begin{figure}[hbt]
\centering
\includegraphics[width=0.7\textwidth]{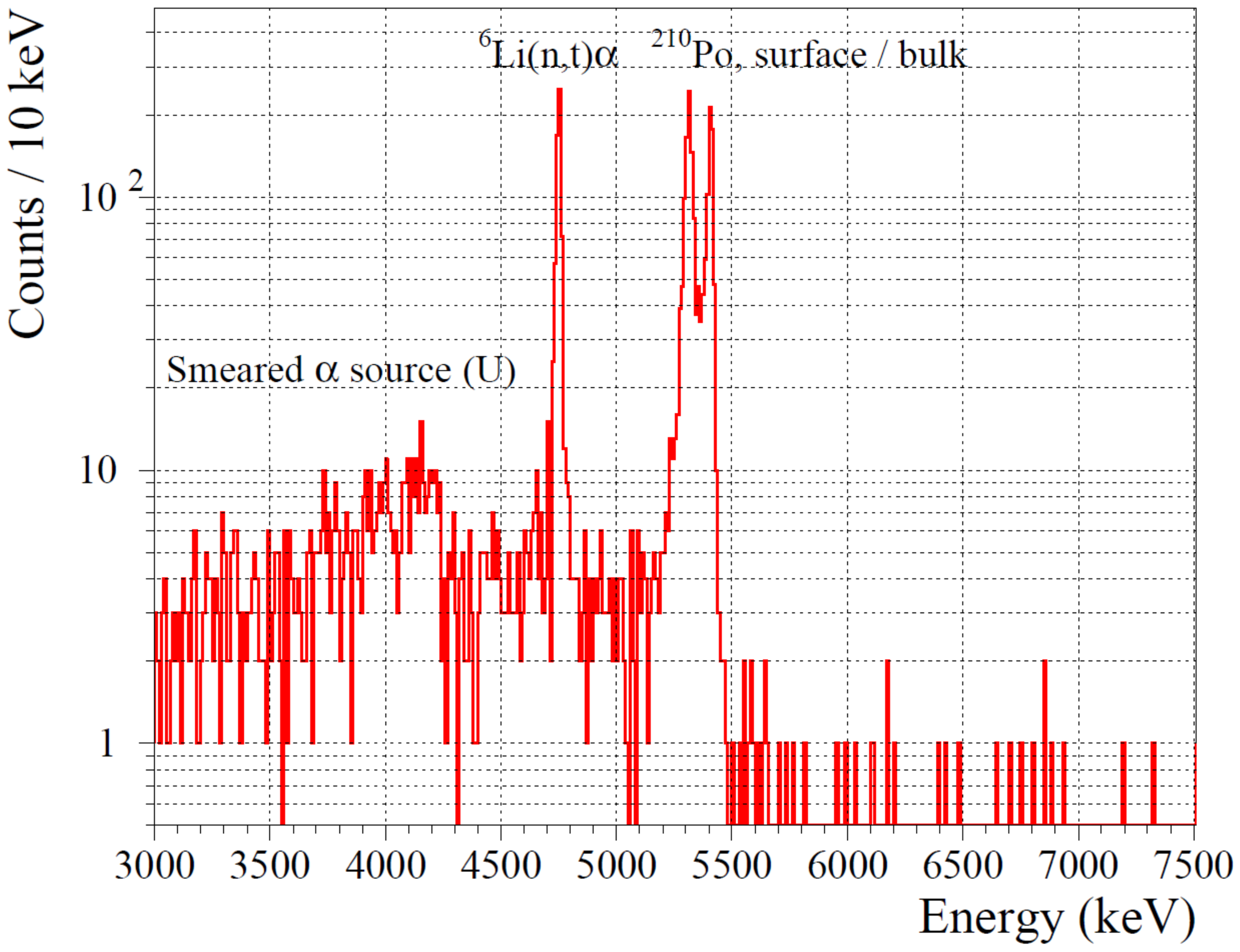}
\caption{Energy spectrum of $\alpha$ events acquired by five 0.28 kg LMO bolometers over 64 days of measurements in the CROSS cryogenic facility at LSC.}
\label{fig:LMO_Bkg_alpha}
\end{figure}

In order to investigate radioactivity induced by U/Th contaminants in LMOs crystal bulk we used background data acquired with the five best-performing enriched LMOs (according to table \ref{tab:LMO_performance}). We selected data characterized by a long duration (>200 h), steady temperature and noise conditions, amounting to 64 days of measurements (89 kg$\times$d exposure) in total. The combined energy spectrum of $\alpha$ particles detected by the five LMO detectors is shown in figure \ref{fig:LMO_Bkg_alpha}. 

Several features seen in figure \ref{fig:LMO_Bkg_alpha} are common for all five detectors. The energy spectra of $\alpha$ particles detected by LMOs with energies below 4.7 MeV are dominated by events induced by the smeared $^{238}$U/$^{234}$U $\alpha$ sources placed in front of the Al-coated LMOs. 

A sharp peak at around 4.8 MeV is originated to the detection of products of thermal neutron captures by $^6$Li, leading to the $^6$Li(n,t)$\alpha$ reaction. 
The counting rate of this peak is consistent among all detectors, and is about 1.8(2)~counts/day. 

A doublet peaked at 5.3 and 5.4 MeV is caused by $\alpha$ decays of $^{210}$Po on the detector surface and crystal bulk, respectively. The 0.1 MeV difference is explained by the detection of $^{206}$Pb nuclear recoils in the later case. The activity of $^{210}$Po is found to be in the range of 40--150~$\mu$Bq/kg, 62~$\mu$Bq/kg in average. A similar variation of the $^{210}$Po activity is observed for LUMINEU \cite{Armengaud:2017,Poda:2017a} and CUPID-Mo \cite{Armengaud:2020a,Poda:2020} crystals; it is mainly originated to the $^{210}$Pb contamination typical for crystal scintillators, particularly for molybdates \cite{Armengaud:2015,Armengaud:2017,Danevich:2017,Danevich:2018}. 
Note we present here a very conservative estimate of the $^{210}$Po bulk activity, because it includes a possible contribution from the crystal surface $^{210}$Po contamination; in order to disentangle this contribution, a more sophisticated analysis of the background model is required \cite{Azzolini:2019nmi,CUPIDMoBkgModel:2023}.  

The sensitivity to detect peaks of $^{232}$Th (4.1 MeV), $^{238}$U (4.3 MeV), and $^{226}$Ra (4.8 MeV) is strongly affected by the $\alpha$ source and neutron-induced backgrounds, thus we searched for decays of $^{232}$Th and $^{226}$Ra $\alpha$-decaying daughters with Q-values within the 5.5--7.5 MeV energy interval, in the region with a rather low counting rate of 0.43(7)~counts/day/kg. In this energy range we see no clear evidence of the presence of peaks of $^{228}$Th (5.5 MeV) and $^{224}$Ra (5.8 MeV) from the $^{232}$Th chain and/or $^{222}$Rn (5.6 MeV) and $^{218}$Po (6.1 MeV) from the $^{238}$U family. Thus, we can set only upper limits on activities of $^{228}$Th and $^{226}$Ra contaminants in the bulk of enriched LMOs:  below 0.7 and 1.1~$\mu$Bq/kg, respectively. Such high radiopurity of LMOs is ensured by the LUMINEU technology of crystal production from additionally purified $^{100}$MoO$_3$, which satisfies CROSS and CUPID demands on the ultra-low level of radioactive contaminants in high-quality LMO crystals.

\section{Conclusions}

We realized a 12-crystal array of LMO scintillating bolometers, accompanied by 16 Ge bolometric light detectors, to test different detector configurations for the future double-beta decay experiments CROSS and CUPID. We installed the array in the CROSS cryogenic facility at the Canfranc Underground Laboratory, cooled down to 14 mK, and operated over almost a half of year with about 80\% duty cycle of data taking. Despite the sub-optimal noise conditions ---significant pulse-tube induced noise, in the detector signal bandwidth, present for most of the channels--- we achieved satisfactory results in terms of detector performance, which allowed us to investigate crucial parameters for the design of CROSS and CUPID detector structures (i.e. performance of detectors instrumented with different sizes of phonon sensors and in different scintillation light collection conditions).

We found no clear difference in performance of LMO bolometers equipped with NTD thermistors with a factor of 3 different volume (3$\times$3$\times$1 mm and 9$\times$3$\times$1 mm), but a larger voltage signal (larger NTD resistance) and a hint on a few ms shorter rise time of LMOs instrumented with larger NTDs. On the contrary, all 3 operational LMO bolometers with four lateral surfaces being covered with an Al layer (2.0 $\mu$m or 0.2 $\mu$m thick) show a reduced sensitivity and notably faster time response. Even if noise conditions were far from optimal, the LMO detectors are characterized by reasonably good energy resolution, in particular the resolution for the 2615 keV gamma peak of $^{208}$Tl in the combined spectrum of the five best performance LMOs was measured as 7.7(2) keV FWHM. Radioactive contaminants from U/Th chains were investigated by analysis of alpha peaks in five enriched LMO crystals. The activities of $^{228}$Th and $^{226}$Ra are found to be below 1 $\mu$Bq/kg, which satisfies demands of CROSS and CUPID on crystal bulk radiopurity.

Most of Ge LDs (12 over 16) were strongly affected by the pulse-tube induced noise, consequently their performance is also impacted by stronger NTD currents applied to mitigate the issue. However, neither the position in the tower nor the NTD size are expected to be responsible for the observed difference in the LD performance, while the tower suspension is more likely to be responsible for the noise issue. Indeed, the tower design is identical for three LMO floors sandwiched by four LD floors. A similar level of LD sensitivity at the same operational point (i.e. NTD current) was observed by analysis of the LDs response to light signals produced by LED. In contrast to a significant difference in the LDs noise, which varies from 0.06 to 0.46 keV FWHM, the LD energy resolution for 17.5 keV Mo X-ray peaks is found to be similar, on the level of 1 keV FWHM.

Despite the noise excess of LMOs and LDs, we were able to study the LMO scintillation detection in various light collection conditions. We observed in average (20--30)\% higher photons collection with square LDs compared to circular ones. The highest light collection was observed for LMOs surrounded by a reflective film; the square and circular LDs measured an average scintillation signal of 0.60(4) and 0.49(2) keV per 1 MeV energy deposited in the LMO crystal scintillator. The absence of the reflector around LMO scintillator reduces the light collection efficiency by about a factor 2, to an average signal of 0.24(4) and 0.18(2) keV/MeV, respectively. These results are similar to values reported in recent studies with cylindrical and cubic LMO scintillating bolometers. 
Surprisingly, even lower scintillation signal is detected from the Al-coated LMOs, in average 0.15(3) and 0.11(2) keV/MeV for square and circular LD, respectively. Taking into account the poor bolometric and scintillation signals, the present procedure of Al coating of four lateral surfaces of LMOs cannot be adopted for CROSS and CUPID. However, some features of coated detectors, like a fast response and ability to identify surface events \cite{Bandac:2021}, motivate us to R\&D further the coating procedures to achieve the ultimate goal.

\acknowledgments

The CROSS and CUPID Collaborations thank the directors and staff of the Laboratorio Subterr\'aneo de Canfranc and the technical staff of our laboratories. This work was supported by the Istituto Nazionale di Fisica Nucleare (INFN); by the European Research Council (ERC) under the European Union Horizon 2020 program (H2020/2014-2020) with the ERC Advanced Grant no. 742345 (ERC-2016-ADG, project CROSS) and the Marie Sklodowska-Curie Grant Agreement No. 754496; by the Italian Ministry
of University and Research (MIUR) through the grant Progetti di ricerca di Rilevante Interesse Nazionale (PRIN 2017, grant no. 2017FJZMCJ); by the US National Science Foundation under Grant
Nos. NSF-PHY-1614611 and NSF-PHY-1401832;  
by the P2IO LabEx (ANR-10-LABX-0038) managed by the Agence Nationale de la Recherche and by the IN2P3 (project CUPID-TDR) in France. 
This material is also based upon work supported by the US Department of Energy (DOE) Office of
Science under Contract Nos. DE-AC02-05CH11231 and DE-AC02-06CH11357; and by the DOE Office of Science, Office of Nuclear Physics under Contract Nos. DE-FG02-08ER41551, DE-SC0011091, DE-SC0012654, DE-SC0019316, DE-SC0019368, and DE-SC0020423. This work was also supported by the National Research Foundation of Ukraine under Grant No. 2020.02/0011 and by the National Academy of Sciences of Ukraine in the framework of the project ``Development of bolometric experiments for the search for double beta decay'', the grant number 0121U111684. \\
 More information on the technical details of the CUPID experiment and the collaboration policies can be found at the official web-page (\href{https://cupid.lngs.infn.it}{https://cupid-it.lngs.infn.it}).


\end{document}